\documentclass{appolb}
\usepackage{graphicx}

\usepackage{latexsym} 
\usepackage{amssymb} 
\usepackage{color}  
\usepackage{amsmath}
\usepackage{hyperref} 
\newcommand{\eq}{{\rm eq}}
\newcommand{\taueq}{\tau_R}
\newcommand{\pres}{{\cal P}}
\newcommand{\ped}{{\cal E}}
\newcommand{\pedeq}{\ped_0}
\newcommand{\preseq}{{\pres_0}}
\newcommand{\deltaf}{\delta\! f}
\newcommand{\deltaB}{\delta\! B}

\def\dM{\kappa}

\newcommand{\beq}{\begin{equation}}
\newcommand{\eeq}{\end{equation}}
\newcommand{\bea}{\begin{eqnarray}}
\newcommand{\eea}{\end{eqnarray}}

\begin{document}
\title{
Thermodynamically consistent formulation\\ of quasiparticle viscous hydrodynamics
\thanks{Presented at
Excited QCD 2017, 7-13 May 2017, Sintra, Portugal.}%
}
\author{Radoslaw Ryblewski
\address{Institute of Nuclear Physics, Polish Academy of Sciences, PL-31342 Krak\'ow, Poland}
}
\maketitle
\begin{abstract}
A novel formulation of second-order relativistic viscous fluid dynamics based on the effective Boltzmann equation for quasi-particles with medium-dependent masses is briefly reviewed.~The evolution equations for the shear and bulk dissipative corrections, and the corresponding transport coefficients, are presented. Resulting approach allows for thermodynamically-consistent incorporation of lattice QCD equation of state in the fluid dynamical framework.
\end{abstract}
\PACS{12.38.Mh, 25.75.-q, 47.75.+f, 12.39.Ba}
%
%%%%%%%%%%%%%%%%%%%%%%%%%%%%%%%%%%%%%%%%%%%%%%%%%% 
\section{Introduction}
%%%%%%%%%%%%%%%%%%%%%%%%%%%%%%%%%%%%%%%%%%%%%%%%%%
%
One of the main goals of the ultra-relativistic heavy-ion collision experiments is to study strongly-interacting QCD matter at extreme conditions. It is believed that such a hot and dense system will reach proximity of the local thermal equilibrium (LTE) allowing to study thermodynamic and transport properties of the system, as encoded in its equation of state (EoS) and kinetic coefficients. If LTE is achieved immediately after collision, and the interactions maintain this state during subsequent evolution, the fluid dynamical description of the system may be applicable.~Successful description of the experimental data on correlations and fluctuations within fluid dynamical framework proves adequacy of such an approach \cite{Florkowski:2010zz,Jaiswal:2016hex,Florkowski:2017olj}.~Despite its successes, fluid dynamics is still plagued with a number of problems.~Of particular importance is the question of a thermodynamically-consistent incorporation of the realistic EoS \cite{Florkowski:2017olj}, such as the one obtained in lattice QCD (lQCD) calculations.

~Canonically, fluid dynamics is constructed as an order-by-order expansion around equilibrium state in powers of thermodynamic gradients, where the zeroth-order theory is that of ideal fluid dynamics \cite{Florkowski:2010zz,Jaiswal:2016hex,Florkowski:2017olj}.~The universal existence of dissipative effects in all physical systems and requirement of causality implies usage of viscous fluid dynamics at least at its second order.~Unlike the non-relativistic first-order Navier-Stokes theory the form of the relativistic fluid dynamical equations of motion is not universal.~Usually, a simple relativistic kinetic theory of an ideal gas of uncharged on-shell particles following the Boltzmann equation, is used to construct equations of motion \cite{Florkowski:2010zz,Jaiswal:2016hex,Florkowski:2017olj}. In such a case, the problem arises how to incorporate a realistic EoS in a thermodynamically-consistent way \cite{Gorenstein:1995vm}.  

In this proceedings contribution we first explain the problem of breaking of the basic thermodynamic relations in the relativistic kinetic theory description of quasi-particles with medium-dependent masses.~Subsequently, following  Ref.~\cite{Tinti:2016bav}, we present a method to derive the second-order viscous fluid dynamical equations and respective transport coefficients based on introducing the idea of the non-equilibrium (bag) mean-field and using the \emph{effective} Boltzmann equation. Resulting approach allows for introducing in a thermodynamically-consistent way any realistic EoS.

%%%%%%%%%%%%%%%%%%%%%%%%%%%%%%%%%%%%%%%%%%%%%%%%%%
\section{Quasi-particles and thermodynamic consistency}
%%%%%%%%%%%%%%%%%%%%%%%%%%%%%%%%%%%%%%%%%%%%%%%%%%
When deriving relativistic fluid dynamics from kinetic theory one usually considers the system of ideal (non-interacting) uncharged on-shell particles of a single species. The resulting EoS of such a system depends only parametrically on the mass of the constituents ($m={\rm const.}$) \cite{Florkowski:2010zz}. Hence, within this approach it is impossible to reproduce a realistic EoS, such as the one obtained using lQCD calculations. 

A possible solution of this problem is to introduce notion of quasi-particles and consider temperature dependence of (now treated as \emph{effective}) mass, $m=m(T)$ \cite{Goloviznin:1992ws}.~One should note that while at large-$T$ limit of QCD the idea of quasi-particles may be physically sound ($m\sim T$) \cite{Braaten:1989mz}, at low temperatures (especially in the QCD phase-transition region), the quasi-particles do not correspond to any real excitations of the underlying theory. The procedure quoted above has a major drawback of violating the basic thermodynamic relations \cite{Gorenstein:1995vm}. The thermodynamic consistency may be restored by
introducing additional mean field $B_0(T)$, whose presence gives rise to the effective in-medium masses \cite{Gorenstein:1995vm}. The $B_0(T)$ field may be included in the Lorentz covariant way by modifying the kinetic-theory definition of the equilibrium energy-momentum tensor \cite{Jeon:1994if,Jeon:1995zm,Chakraborty:2010fr,Romatschke:2011qp,Albright:2015fpa}
\begin{equation}
T_\eq^{\mu\nu} = \int dP \, p^\mu p^\nu \, f_\eq   + B_0(T) \, g^{\mu\nu},  
\label{eqbag}
\end{equation}
where $\int dP  =  \int d^4 p \, 2 \, \Theta(p^\mu t_\mu)  \, \delta(p^\mu p_\mu - m^2)/ (2\pi)^3$ (with $t_\mu$ being arbitrary time-like vector), $f_\eq$ is the equilibrium distribution function, and $g^{\mu\nu}={\rm diag}\left(1, -1,-1,-1\right)$ is the metric tensor. Using definitions of energy density and pressure 
\begin{equation}\label{eng_prs}
\pedeq = u_\mu T_\eq^{\mu\nu}   u_\nu, \qquad \preseq = -\frac{1}{3}\Delta_{\mu\nu} T_\eq^{\mu\nu},
\end{equation}
respectively, where $u^\mu$ is the four-velocity of the fluid in the Landau frame, and $\Delta^{\mu\nu}=u^\mu u^\nu -g^{\mu\nu}$,  one finds that satisfying thermodynamic relation
\begin{equation}\label{th_rel}
{\cal 
S}_0\equiv\frac{d\preseq}{dT} = \frac{\pedeq + \preseq}{T},
\end{equation}
requires 
\begin{equation}\label{th_rel}
\frac{dB_0}{dT} =- m \, \frac{dm}{dT}\int dP f_\eq.
\end{equation}
We note here, that the entropy density ${\cal 
S}_0$ in Eq.~(\ref{th_rel}) is independent of the $B_0(T)$ field, which suggests that it may be used to extract $m(T)$ from the lQCD data \cite{Romatschke:2011qp}.
%
%%%%%%%%%%%%%%%%%%%%%%%%%%%%%%%%%%%%%%%%%%%%%%%%%%
\section{Non-equilibrium mean field}
%%%%%%%%%%%%%%%%%%%%%%%%%%%%%%%%%%%%%%%%%%%%%%%%%%

In the case when the system is out of equilibrium,  Eq.~(\ref{eqbag}) has to be generalized to \cite{Tinti:2016bav}
\begin{equation} 
 T^{\mu\nu} = \int dP \, p^\mu p^\nu \, f  \;  + B^{\mu\nu},
\label{noneqbag}
\end{equation}
where the distribution function  now contains the non-equilibrium correction, $f=f_\eq+\deltaf$. It is thus natural to split the second term in Eq.~(\ref{noneqbag}) into equilibrium and non-equilibrium part, $B^{\mu\nu}=B_0 g^{\mu\nu} + \deltaB^{\mu\nu}$ as well. The non-equilibrium part $\deltaB^{\mu\nu}$ is fixed by the requirement of conservation of energy and momentum, $\partial_\mu T^{\mu\nu}=0$. In general, due to the symmetry of $T^{\mu\nu}$, the non-equilibrium part $\deltaB^{\mu\nu}$ can have ten independent components. However, the energy-momentum conservation provides only four constraints. For that reason we make the following \emph{ansatz}, which restricts the number of independent components of $\deltaB^{\mu\nu}$ \cite{Tinti:2016bav}
\begin{equation}\label{ansatz}
 \deltaB^{\mu\nu} = b_0 \,  g^{\mu\nu} + u^\mu b^\nu + b^\mu u ^\nu,
\end{equation}
where $b^\mu$ satisfies the requirement \mbox{$u_\mu  b^\mu=0$}.

We note here that the notion of the non-equilibrium mean-field was first introduced to include realistic EoS in the quasi-particle formulation of the anisotropic hydrodynamics \cite{Alqahtani:2016rth,Alqahtani:2017jwl,Alqahtani:2017tnq}.
%
%%%%%%%%%%%%%%%%%%%%%%%%%%%%%%%%%%%%%%%%%%%%%%%%%%
\section{Evolution equations for dissipative fluxes}
%%%%%%%%%%%%%%%%%%%%%%%%%%%%%%%%%%%%%%%%%%%%%%%%%%
%
Using definitions of the energy-momentum tensors (\ref{eqbag}) and (\ref{noneqbag}) and the ansatz (\ref{ansatz})
the  bulk pressure $\Pi$ and  shear tensor $\pi^{\mu\nu}$ corrections are given by
\begin{align}
 \Pi &\equiv -\frac{1}{3}\Delta_{\alpha\beta}\left( T^{\alpha\beta}  - T_\eq^{\alpha\beta} \right) =-\frac{1}{3}\Delta_{\alpha\beta} \int dP p^\alpha p^\beta \deltaf - b_0, \label{bulk_def}\\
 \pi^{\mu\nu} &\equiv \Delta^{\mu\nu}_{\alpha\beta} \left( T^{\alpha\beta} 
 - T^{\alpha\beta}_\eq \right) = \Delta^{\mu\nu}_{\alpha\beta} \int dP p^\alpha p^\beta \deltaf \label{shear_def},
\end{align}
respectively, where $\Delta^{\mu\nu}_{\alpha\beta}=\frac{1}{2} 
(\Delta^{\mu}_{\alpha}\Delta^{\nu}_{\beta} + 
\Delta^{\mu}_{\beta}\Delta^{\nu}_{\alpha} - \frac{2}{3} 
\Delta^{\mu\nu}\Delta_{\alpha\beta})$ is the symmetric traceless 
projector orthogonal to $u^\mu$.
The equations of motion for the dissipative quantities may be obtained using the Chapman-Enskog-like iterative solution \cite{Jaiswal:2013npa,Jaiswal:2013vta,Bhalerao:2013pza,Jaiswal:2014isa} of the effective Boltzmann equation for quasi-particles with $T$-dependent mass \cite{Romatschke:2011qp,Florkowski:1995ei} 
\begin{equation}\label{BE}
 p_\mu \partial^\mu f + m\, (\partial^\rho m) \, \partial_\rho^{(p)}\, f  = {\cal C}[f].
\end{equation}
For the sake of simplicity we treat the collision term in the relaxation-time approximation \cite{Anderson:1974}
\begin{equation}\label{RTA}
 {\cal C}[f] = -\frac{p_\mu u^\mu}{\taueq}\deltaf,
\end{equation}
where $\taueq$ is the relaxation time.

At first order in gradients one gets \cite{Tinti:2016bav,Romatschke:2011qp}
\begin{equation}\label{navier_stokes}
 \Pi = - \beta_{\Pi} \, \taueq \, \theta, \quad \pi^{\mu\nu} = 2 \, \beta_\pi \taueq \, \sigma^{\mu\nu},
\end{equation}
with 
\begin{align}
 \beta_\Pi =\;& \frac{5}{3}\, \beta \, I_{3,2} - c_s^2\left( \ped + \pres \right) 
 + \dM c_s^2 \, m^2 \beta \, I_{1,1} ~,\label{beta_Pi}\\
 \beta_\pi =\;& \beta I_{3,2}. \label{beta_pi}
\end{align}
In Eqs.~(\ref{navier_stokes})-(\ref{beta_pi}) we introduced $\beta\equiv1/T$, speed of sound squared $c_s^2=d\pres/d\ped$, $\dM\equiv(T/m)(dm/dT)$, stress tensor $\sigma^{\mu\nu}\equiv\Delta^{\mu\nu}_{\alpha\beta} 
\nabla^\alpha u^\beta$, $\theta\equiv \partial_\mu u^\mu$, $\nabla^\mu\equiv\Delta^{\mu\nu}\partial_\nu$, and the functions 

\begin{equation}
 I_{n,q} \equiv \frac{(-1)^q}{(2q + 1)!!}\int dP \, (p_\mu u^\mu)^{n-2q}\left(p_\mu \Delta^{\mu\nu}  p_\nu\right)^q f_\eq.
\end{equation}
The shear and bulk  viscosities are given by the following relations: $\beta_\pi \taueq = \eta$ and $\beta_\Pi \taueq = \zeta$. One should stress here that, while the form (\ref{beta_pi}) of shear viscosity is the same as in the case of constant masses  \cite{Jaiswal:2014isa}, the expression for bulk viscosity (\ref{beta_Pi}) contains additional contribution due to in-medium mass. 

Applying the co-moving derivative $\dot{\left(\,\, \right)} \equiv u_\mu \partial^\mu$ to Eqs.~(\ref{bulk_def})-(\ref{shear_def}) \cite{Denicol:2010xn} one arrives at the second-order viscous fluid equations for shear and bulk corrections of the form \cite{Denicol:2010xn,Denicol:2012cn,Denicol:2014mca,Jaiswal:2014isa,Tinti:2016bav}
\begin{align}
\dot{\Pi} =& -\frac{\Pi}{\tau_{\Pi}}
-\beta_{\Pi}\theta 
-\delta_{\Pi\Pi}\Pi\theta
+\lambda_{\Pi\pi}\pi^{\mu\nu} \sigma_{\mu\nu}, \label{BULK}\\
\dot{\pi}^{\langle\mu\nu\rangle} =& -\frac{\pi^{\mu\nu}}{\tau_{\pi}}
+2\beta_{\pi}\sigma^{\mu\nu}
+2\pi_{\gamma}^{\langle\mu}\omega^{\nu\rangle\gamma}
-\tau_{\pi\pi}\pi_{\gamma}^{\langle\mu}\sigma^{\nu\rangle\gamma}  
 -\delta_{\pi\pi}\pi^{\mu\nu}\theta 
+\lambda_{\pi\Pi}\Pi\sigma^{\mu\nu}, \label{SHEAR}
\end{align}
where $\omega^{\mu\nu}\equiv\frac{1}{2}(\nabla^{\mu}u^{\nu}- 
\nabla^{\nu}u^{\mu})$ is the vorticity tensor.

The respective transport coefficients have the following form \cite{Tinti:2016bav}
\begin{align} 
 \delta_{\Pi\Pi} =&\; -\frac{5}{9} \chi - \left( 1 - \dM m^2 
 \frac{I_{1,1}}{I_{3,1}} \right) c_s^2  
  + \frac{1}{3} \frac{\beta \dM c_s^2 m^2 }{\beta_\Pi} \Bigg[
  \left( 1-3c_s^2 \right) \left( \beta I_{2,1} - I_{1,1}\right) \nonumber\\
 & -\left(1  -3\dM c_s^2 \right)
m^2 \left( \beta I_{0,1} + I_{-1,1} \right) \Bigg], \label{coeff1L}\\
  \lambda_{\Pi\pi} =&\; \frac{\beta}{3\beta_\pi} \left( 2 I_{3,2} - 7 I_{3,3}  \right) 
 -  \left( 1 - \dM m^2 
 \frac{I_{1,1}}{I_{3,1}} \right) c_s^2, \label{coeff2L}\\
  \tau_{\pi\pi} =&\; 2 -  \frac{4\beta}{\beta_\pi} \, I_{3,3}, \qquad \lambda_{\pi\Pi} =\; -\frac{2}{3} \chi, \label{coeff3L}\\
   \delta_{\pi\pi} =&\; \frac{5}{3} - \frac{7}{3} \frac{\beta}{\beta_\pi} \, I_{3,3} 
   - \frac{\beta}{\beta_\pi}\, \dM c_s^2 m^2  \left( I_{1,2}-I_{1,1} \right), \label{coeff4L} 
\end{align}
with $\chi \equiv \beta \left[ \left( 1- 3 c_s^2 \right) 
 \left(  I_{3,2}  -I_{3,1} \right)  
 -  \left(1  -3\dM c_s^2 \right)
m^2 \left(I_{1,2}- I_{1,1}\right)\right]/\beta_\Pi$. As expected, in the massless limit Eqs.~(\ref{BULK})-(\ref{coeff4L}) reduce to the ones from Ref.~\cite{Jaiswal:2014isa}. When supplemented with the four equations of motion for the energy density and four-velocity, and the realistic EoS (through the form of $m(T)$ and $B_0(T)$) they provide the thermodynamically-consistent framework to describe evolution of matter in the heavy-ion collisions, see Ref.~\cite{Tinti:2016bav}.
%
%%%%%%%%%%%%%%%%%%%%%%%%%%%%%%%%%%%%%%%%%%%%%%%%%%
\section{Summary}
%%%%%%%%%%%%%%%%%%%%%%%%%%%%%%%%%%%%%%%%%%%%%%%%%%
%
In this proceedings contribution we shortly reviewed a new formulation of second-order relativistic fluid dynamics for the system made of  quasi-particles with medium-dependent masses based on the effective Boltzmann equation \cite{Tinti:2016bav}. Unlike other hydrodynamical models, the presented approach provides first hydrodynamical framework, which allows for introducing a realistic lQCD-based EoS in a thermodynamically-consistent way. 
%
%%%%%%%%%%%%%%%%%%%%%%%%%%%%%%%%%%%%%%%%%%%%%%%%%%
\section*{Acknowledgments}
%%%%%%%%%%%%%%%%%%%%%%%%%%%%%%%%%%%%%%%%%%%%%%%%%%
%
This work was supported by Polish National Science Center Grant No. DEC-2012/07/D/ST2/02125.
%
%%%%%%%%%%%%%%%%%%%%%%%%%%%%%%%%%%%%%%%%%%%%%%%%%%

\end{document}